\input{aipcheck}

\documentclass[
    ,final            % use final for the camera ready runs
  ]
  {aipproc}

\layoutstyle{8x11double}

\begin{document}

\title{Spin-Phonon Coupling in High-T$_c$ Copper Oxides.}


\author{T. Jarlborg}{
  address={DPMC, University of Geneva, CH1211 Geneva 4, Switzerland}
}

%\author{<author2>} address={<common address for author2 and author3>}
%}

\begin{abstract}
 Band calculations on HgBa$_2$CuO$_4$ and La$_{(2-x)}$Sr$_x$CuO$_4$
 with phonon and spin-waves within the CuO planes show that partial gaps
 are created at various energies depending on wavelengths. Spin and phonon gaps appear
 at different energies when the modulations are along [1,1,0], while
 they are at the same energy for modulations along [1,0,0]. 
 It is shown that the ability to form gaps and antiferromagnetic waves
 is correlated with the strength of the interaction parameter $\lambda_{sf}$
 for spin fluctuations. Many unusual properties of the high-T$_C$ oxides
 can be understood from spin-phonon coupling.

\end{abstract}

\maketitle

%%%%%%%%%%%%%%%%%%%%%%%%%%%%%%%%%%%%%%%%%%%%
%% MAINMATTER
%%%%%%%%%%%%%%%%%%%%%%%%%%%%%%%%%%%%%%%%%%%%

\section{Introduction}
Properties of
high-$T_C$ superconductors have been seen to
depend both on phonons \cite{iso,phon} and magnetic 
fluctuations \cite{mag,mok}.
Therefore it is expected that the mechanism for high-$T_C$ 
superconductivity should involve phonons as well as
magnetic fluctuations. Here, band results show that spin-phonon
coupling (SPC) is important and many results are compatible
with observations.

%\section{<Another section>}

Band calculations on the high-$T_C$ cuprates HgBa$_2$CuO$_4$ and 
La$_{(2-x)}$Sr$_x$CuO$_4$ show that gaps or pseudo gaps appear
near the Fermi energy, $E_F$, when the potential has a modulation
within the CuO-plane \cite{jar1}. The calculations are made for
long unit cells oriented either along [1,1,0] or [1,0,0] containing so-called
half-breathing (bond-stretching) phonons. The planar O-atoms on both sides of
a Cu-atom are alternatively displaced towards ("compressed") or
away from the Cu ("diluted"). Calculations with anti-ferromagnetic (AFM)
field on the Cu atoms are made to model spin-waves. 
When the calculations are made for
co-existing phonon and spin waves, it turns out that the
magnetic moments are largest on the "diluted" Cu-sites.
The nodes of modulated spin waves coincide with "compressed"
Cu-sites. 
The phonon mode
creates a potential modulation of the potential (a Fourier component
$V_Q$). Spin waves
induce a modulation in the spin-polarized part of the potential
and magnetic moments on the Cu. Gaps appear at different energies
depending on the wave length of the modulation. The gain in 
kinetic energy is maximized when the gap is at $E_F$.
The position of $E_F$ is controlled by doping
(calculations
use the virtual crystal approximation).
The wavelength of the spin wave is twice that of the phonon,
because the spin can be up or down at "diluted" Cu-positions. Hence,
for modulations along [1,1,0],
the band structure has two gaps, one due to the phonon and
one at a higher energy because of the spin wave, and there is no direct SPC
for electrons at $E_F$. However, 
for modulations along [1,0,0]
one realizes that two rows of CuO are required along the unit cell, 
so that the phase of the spin wave along each row will
differ by $\pi$. 
The band results show in this case a constructive SPC,
where the gap from separate phonon and AFM-modulations will open a gap
at the same energy, i.e. at $E_F$ for the correct doping. 
In addition, a phonon with large atomic displacements
will increase the moments of the spin wave, and both waves contribute to
a common gap.

 The abovementionned results can be found in refs, \cite{jar1}-\cite{jar4},
 where it also is shown that a small correction to the band structure (which
 affects the potential and the localization) will stabilize
a gapped AFM state instead of the metallic non-magnetic state for the undoped
system. The same correction will 
increase the mass enhancement from spin fluctuations in doped cases \cite{jar4}.
%Here, models of the band results are used to estimate some key properties.
 
\section{Results and discussion}

The calculated density of states (DOS)
 at $E_F$, 
$N \approx 1.5 (Cu \cdot eV)^{-1}$. 
The atomic displacement $u^2=1.5\hbar \omega/K$ at low $T$ and $u^2=3k_BT/K$ 
at high $T$ (relative to the Debye temperature).
The force constant $K=25$ eV/\AA$^2$ for a typical O phonon with 
$\hbar \omega \sim$ 80 meV makes $u$ 
about 0.06 \AA ~at room temperature. Spin waves can, in analogy with phonons,
be assigned a magnetic moment $m^2= k_BT/K_m$, where $K_m = d^2E_{tot}/dm^2$.
This gives $m \approx$ 0.3 $\mu_B$/Cu at room temperature for the co-existing case.
A partial gap removes roughly half of the DOS 
within an energy $\Delta =$ 100 meV around $E_F$. The gain in kinetic energy
is approximately $\frac{1}{2} N \Delta^2 =  8$ meV/Cu, i.e. about 15-20 percent
of the elastic energy $U =\frac{1}{2}K \cdot u^2$ at the average $u$. Thus, the phonon
energy is expected to decrease by the this percentage because of the coupling
to the spin wave, but only in the doped case when $E_F$ is at the gap. These
results compare resonably well with experiment \cite{phon}. Less doping implies
coupling at longer wavelengths so that the softening moves away from the zone
boundary and finally disappears. The softening should also disappear when the
pseudo gap disappears. This is expected when the Fermi-Dirac occupation, $f$,
becomes too wide to separate the occupied majority spins at $E_F - \Delta$
from the unoccupied minority spins at $E_F + \Delta$. 
As function of $T$,
$m(T)= N \cdot \Delta \cdot (f(\frac{-\Delta}{k_BT})-f(\frac{\Delta}{k_BT})$,
while from the band calculations $\Delta \approx const. \cdot m$. By solving iteratively
the two last equations as function of $T$, one finds a stable gap
(and $m$) up to a $0.9 \cdot T_{max}$ after which $\Delta$ and $m$ drop to zero
at $T_{max}$, as is the case with the pseudo gap near $T^*$. The reason
is the strong exponential $T$-dependence of $f$, and the feedback of $m$ on $\Delta$.

Isotope shifts are expected. The zero-point motion makes a mass ($M$)
dependence $u \sim (K \cdot M)^{-1/4}$. A lower $M$ will also decrease
$K$ because of the phonon softening due to the promotion of spin waves
at larger u (see above), 
so the isotope effect will be stronger than what is suggested from a
constant $K$. The effect on the pseudogap will be moderated at large T,
when there is no explicit mass dependence of $u$, $u^2 \sim K^{-1}$.
The effect on the superconducting $T_c \sim \omega exp(-1/\lambda)$ is complex.
Pure phonon $\omega \sim \sqrt{K/M}$, but with SPC also
$K$ will drop when phonon amplitudes increase because of lower $M$.
A pure $\lambda_{sf}$ from spin fluctuations tend to increase at increased
phonon amplitudes. The mass dependence of $\omega_{sf}$ and $\lambda_{sf}$ may act
oppositely on $T_c$. Anharmonic effects are expected, which will mix phonon
and spin contribution. Experiments show small isotope effects on $T_c$ \cite{iso}.

The mass dependence of $u$ is largest at low $T$, which suggests comparison
with isotope shifts of the penetration depth $\Lambda$ measured at low $T$ \cite{iso}.
Some assumptions lead to $\Lambda^{-2} \sim N \cdot v^2$, 
where $v$ is the Fermi velocity. The band crossing $E_F$
along $\Gamma$-M has larger $v$ than at the crossing
along X-M. Phonon and spin waves along [1,0,0] will produce the gaps
in the latter region, so that $v$ will increase. If O$^{16}$ are substituted by
O$^{18}$ it will decrease $u$ for half breathing phonons by 3 percent. The band
results tell that the DOS is reduced roughly by a factor $\frac{1}{2}$ by spin fluctuations
if a certain phonon distortion is present and by 20-25 percent if no distortions are present.
By simple interpolation from these results one can estimate that reducing $u$ by 3 percent
will increase $N$ by 1-2 percent. As $N \sim v^{-1}$ this suggests a decrease
of $\Lambda^{-2}$ by 1-2 percent. This is smaller but of the same order as in ref \cite{iso}.
It can be noted that the gap starts in the X-M region from waves along [1,0,0] (and [0,1,0]).
This fits to the observation of 'destruction' of the Fermi surface in 
this region of k-space \cite{nor},
while it remains along $\Gamma$-M if no waves are oriented along [1,1,0].

Photoemission and tunneling spectra show the pseudogap and a 'dip' in the DOS
about a tenth of an eV below $E_F$. The position of this dip relative to the main
gap at $E_F$ depends on
doping according to tunneling data \cite{Zas}. A projection of these data on
to the calculated DOS suggests that the dip
is located at a fixed energy (a fixed wavelength in our mechanism),
while the main gap moves to higher energy for decreased doping.
This is consistent with a type of perpendicular spin-phonon coupling
as follows: Consider the shortest possible co-existing 
SPC along [1,0,0]. The half-breathing phonon
creates maximum 'diluted' distortions for two out of four pairs of Cu,
where condition for magnetism is optimal (while magnetic nodes fall on the
two other pairs of Cu). The perpendicular rows of (optimally magnetized) Cu can develop
spin waves along [0,1,0], possibly coupled with phonons along the same direction.
The distortion along [1,0,0] has increased the ability for
having strong spin fluctuations along [0,1,0]. While the extent of the
former wave is fixed to 4 cells, it is possible to form longer waves
for the latter so that the main gap is tied to $E_F$ for the given doping.
This perpendicular SPC is expected to have the same properties as discussed above for parallel
coupling. Complete band calculations are more difficult because very
large unit cells are needed to cover even the smallest configuration. 
However, band results for a simplified case, 
with only a half-breathing mode with large and small
moments (no full spin wave) along [1,0,0] together with longer complete
spin waves along [0,1,0], show that two gaps form simultanously at two
different energies. At low evergy the gap is due to the short wave along
[1,0,0], while at larger energy a second partial gap develops because
of a longer spin wave along [0,1,0]. The doping dependence of the two gaps
fits with data in ref. \cite{Zas},
but the energy scale is too large, almost by
a factor of two. This could be explained if the DOS was larger, if all
wavelengths were larger or if the real doping, $x$, is not what it is
supposed to be.

In conclusion, it is shown that many properties of 
high-$T_c$ materials are consistent with SPC, although more
precise works are needed for quantitative
results. More detailed results
will be published elsewhere \cite{jar5}.

\bibliographystyle{aipproc}   % if natbib is available

\bibliography{lt24sample}

\IfFileExists{\jobname.bbl}{}
 {\typeout{}
  \typeout{******************************************}
  \typeout{** Please run "bibtex \jobname" to optain}
  \typeout{** the bibliography and then re-run LaTeX}
  \typeout{** twice to fix the references!}
  \typeout{******************************************}
  \typeout{}
 }

\end{document}